\documentclass[prb,12pt]{revtex4}

\begin{document}
\draft
\title{A new class of nodal stationary states in 2D Heisenberg ferromagnet. }
\author{I.G. Bostrem},
\author{A.S. Ovchinnikov}
\address{Department of Theoretical Physics, Ural State University,
620083, Lenin Ave. 51, Ekaterinburg, Russia}

\date{\today}

\begin{abstract}
 A new class of nodal topological excitations in a two-dimensional
Heisenberg model is studied. The solutions correspond to a nodal
singular point of the gradient field of the azimuthal angle. An
analytical solution found  for the isotropic case. An effect of
in-plane exchange anisotropy is studied numerically. It results in
solutions which are analogues of the conventional  out-of-plane
solitons in the two-dimensional magnets.
\end{abstract}


\maketitle

\newpage
The known topological solutions in two-dimensional (2D) Heisenberg model
belong to classes of the homotopic groups isomorphic to the group of
integers such as $\pi _{2}(S^{2})$: Belavin-Polyakov (BP) solitons \cite
{Belavin} and easy-axis solitons \cite{Kosevich}; to the relative homotopic
group $\pi _{2}(S^{2},S^{1})$: an out-of-plane (OP) \cite{OUTPLAN} and
Takeno-Homma (TH) \cite{Takeno} solitons; and to the group $\pi _{1}(S^{1})$%
: Kosterlitz-Thouless (KT) vortices. They correspond to a map of a spin
order parameter space onto a sphere $S^{2}$ homeomorphic to the 2D plane or
a circle $S^{1}$\cite{Mermin}.

In this paper we use an another way of a search of non-trivial topological
excitations. By starting from the non-linear equations for ($\theta ,\varphi
$)-fields describing the dynamics of a classical 2D isotropic Heisenberg
magnet of a spin $S$ with an exchange $J$ \cite{Bishop}
\begin{equation}
0=-\hbar S\sin \theta \frac{\partial \phi }{\partial t}+JS^{2}\left( \Delta
\theta -\cos \theta \sin \theta \left( \vec{\nabla}\phi \right) ^{2}\right) ,
\label{first}
\end{equation}
\begin{equation}
0=\hbar S\frac{\partial \theta }{\partial t}+JS^{2}\left( 2\cos \theta
\left( \vec{\nabla}\phi \vec{\nabla}\theta \right) +\sin \theta \Delta \phi
\right)   \label{second}
\end{equation}
one may see that in the stationary case the field variable $\theta $ is
determined by only the gradient field $\vec{\psi}=\vec{\nabla}\phi $. The
Euclidean 2D-plane is homeomorphic to the sphere $S^{2}$ with one punctured
point. According to the Hopf theorem \cite{Dubrovin} Eulerian characteristic
of triangulated surface equals to the sum of indices of singular points of
the vector field on the surface. An Eulerian characteristic of the sphere is
equal $2$. In the simplest case one may suggest that one singular point with
the Poincare index $+1$ is placed in the south pole and the other one at
infinity with the same index corresponds to the punctured point in the north
pole. The space configuration of the $\theta $-field will depend on the kind
of singular point of $\vec{\psi}$: a center, a node or a focus. The center
singularity (Fig. 1a) corresponds to the solution of Eq.(\ref{second}) $%
\Delta \phi =0$ and the particular solution $\phi =q\tan ^{-1}\left( \frac{y%
}{x}\right) $ results in the well-known solitons with an axial symmetry
listed above. One may expect that the focus singularity (Fig. 1b) will
correspond to a spiral spin arrangement \cite{Borisov}.

The investigation of the paper is devoted to the nodal point of the vector
field $\vec{\psi}$ (Fig. 1c), where $\vec{\psi}$ has a maximal (minimal)
value. This means the choice of the following parametrization $\phi =\phi
\left( r\right) $, i.e. the azimuthal angle changes along the radial
direction in the plane.
From the equation (\ref{second}) one obtain
\begin{equation}
\frac{d\varphi }{dr}=\frac{q}{r\sin ^{2}\theta },  \label{phider}
\end{equation}
this determines the radial dependence of the $\varphi $-field
\begin{equation}
\varphi (r)=\varphi _{0}+q\int\limits_{a}^{r}\frac{dr^{^{\prime }}}{%
r^{^{\prime }}\sin ^{2}\theta },  \label{phisol}
\end{equation}
where the notation $a$ is used for the lattice unit; $\varphi _{0}$ is an
initial value. The equation for the $\theta $-angle
\begin{equation}
\triangle \theta -\frac{\cos \theta }{\sin ^{3}\theta }\frac{q^{2}}{r^{2}}=0
\label{Thetaeq}
\end{equation}
may be integrated exactly and results in the scale invariant solution
\begin{equation}
\theta (r)=\cos ^{-1}\left[ \sqrt{1-\left( \frac{q}{Q}\right) ^{2}}\sin
\left( p\log \left[ \left( \frac{r}{R}\right) ^{\left| Q\right| }\right]
\right) \right] ,\;p=\pm 1.  \label{BOsol}
\end{equation}
This presents annuluses divergent logarithmically from the center (Fig. 2)
and looks like a ''target'' with annular domains of magnetization. The
parameter $Q^{2}\geq q^{2}$ governs the amplitude of the oscillations, $R$
is a scale factor, the sign $p$ is a polarity of the solution. The continuum
description is valid just for distances greater than the lattice unit $a$,
the $R$ value determines the boundary value $\theta _{0}=\theta |_{r=a}$. As
is seen from Eq. (\ref{BOsol}) the $\theta $ can not take the values $0$ and
$\pi $. The plane chirality is determined by the sign of parameter $q$; the
corresponding in-plane spin texture is presented in Fig. 3.

This solution is a counterpart of the BP soliton because it has no definite
localization radius and it is scale-invariant. In contrary to the BP soliton
the energy of the found solution
\begin{equation}
E=\frac{JS^{2}}{2}\int\limits_{a}^{L}\left[ \left( \vec{\nabla}\theta
\right) ^{2}+\frac{q^{2}}{r^{2}\sin ^{2}\theta }\right] d\vec{r}=\pi
JS^{2}Q^{2}\log \left( \frac{L}{a}\right)   \label{ener}
\end{equation}
has no finite value and reveals the KT logarithmic behavior with an increase
of the system size $L$.

An availability of the pair of stationary solutions resembles in somewhat an exsistence of fundamental system of linear second-order differential equation.
In the study of topological excitations in the classical $XY$ model we deal the last situation: in Ref. \cite{Kosterlitz} just one of the harmonic function $\phi =q\tan ^{-1}\left( \frac{y}{x}\right)$ has been considered as physically reasonable; another solution $\bar \phi = - q \log(r)$ is exploited to obtain an effective interaction between vortices.

As it follows from Eq.(\ref{BOsol}) the choice $\left| Q\right|
=q$ realizes a pure in-plane arrangement $\theta =\frac{\pi
}{2}\,$ and insures a minimal energy $E$ in the class of
solutions.  In the logarithmic scale $x=\log \left(
\frac{r}{R}\right) $ the parameter $Q$ determines the wave length
$\delta =\frac{2\pi }{\left| Q\right| }$, i.e. a distance between
the nearest ''crests'' of the ''target''. The change of the angle
$\varphi $ on the scale $\delta $ is $\triangle \varphi =q\delta $
that leads to the important relation for the small amplitude of
the magnetization oscillations ($q \approx Q$)
\begin{equation}
\frac{\triangle \varphi }{2\pi } \approx \frac{q}{\left| Q\right|
}, \label{kvant}
\end{equation}
i.e. the ratio $q / \left| Q\right| $ responds to the relative
change of the azimuthal angle on the scale $\delta$.

The criterion of topological stability  of the BP soliton is
rather simple  \cite{Kosevich}: there is an integer-valued
topological invariant $Q$ (degree of a map $S^{2} \to S^{2}$)
associated with the BP solution $\theta (\vec r)$, $\phi (\vec r)$
via
$$
    Q= \frac{1}{4 \pi} \int \sin{\theta (\vec r)} d \theta(\vec r) d \phi (\vec r).
$$
For the soliton with an axial symmetry $\phi = \phi (\alpha)$ ($\alpha$ is the angular polar variable) the topological invariant
$$
    Q=\frac{\nu}{2}[\cos \theta(0) -\cos \theta(\infty) ]=\pm \nu,
$$
where $\nu$ is a winding number. This describes a change of the $\phi$ angle at moving around the center $r=0$
$$
    \delta \phi = 2 \pi \nu
$$
and equals to degree of a map $S^{1} \to S^{1}$. A non-zero
density of an angular momenta \cite{Egorov}
$$
    L_{z}=\nu \hbar S (1-\cos \theta)
$$
and a zero value of the radial part of the momentum density
$P_{r}=0$ is a common point of the solitons with the center
singularity: BP and easy-axis solitons, the OP and TH solitons,
and also KT vortices.  Thus, the winding number is associated
uniquely with the angular momenta density of the magnetization
field.

For the solutions with a nodal singularity a situation is
opposite: $L_{z}=0$ and $P_{r}$ has a non-zero value \cite{Egorov}
$$
    P_{r}=\hbar S (1-\cos \theta) \frac{d \phi}{d r} = \hbar \frac{q S}{r} \frac{1}{1+\cos \theta}.
$$
For a pure in-plane spin arrangement the $q$ value  determines a scaling factor
$$
    \log \lambda = \frac{2\pi}{q},
$$
thus, that a change of the azimuthal angle on the scale $\lambda $
$$
    \phi (\lambda r) - \phi (r) = 2 \pi.
$$
As is seen the $q$  value is associated identically with the
momentum density  of the magnetization field.

A special kind of nodal time-dependent solutions of two-dimensional Heisenberg model has been considered in Ref. \cite{Mikhailov} by means of the inverse scattering transform. Keeping time-dependent terms in the Eqs.(\ref{first}, \ref{second}) one may obtain solutions with a finite energy and finite localization radius. The radial behavior of the angle variables differs form  the situation considered above, for example,  the $\theta$  reveals an exponential dependence at large distances in fixed time.  The solitons are ring-shaped waves. Their localization radius and thickness grows with time linearly, whereas an amplitude is inversely proportional to the time.

It is well known that an exchange anisotropy along $z$-axis $J_{z}>J_{\perp }
$ changes the asymptotic behavior at large distances from the power law
decreasing to the exponential ones and results in the easy-axis solitons. An
analogous effect occurs in the considered case but only for the plane
exchange anisotropy $J_{\perp }>J_{z}$. The static equations for the case
with an account of an external magnetic field along $z$-axis $h$ may be
written as
\[
0=J_{\perp }(4\cos \theta \sin \theta +\cos ^{2}\theta \Delta \theta
)-J_{\perp }\cos \theta \sin \theta ((\vec{\nabla}\theta )^{2}+(\vec{\nabla}%
\phi )^{2})
\]
\begin{equation}
-4J_{z}\cos \theta \sin \theta +J_{z}\cos \theta \sin \theta (\vec{\nabla}%
\theta )^{2}+J_{z}\sin ^{2}\theta \Delta \theta -\frac{h}{S}\sin \theta ,
\label{firstani}
\end{equation}
\begin{equation}
0=\sin \theta \Delta \phi +2\cos \theta (\vec{\nabla}\theta \vec{\nabla}\phi
).  \label{secondani}
\end{equation}
The equation (\ref{secondani}) is the same as in the isotropic case and
results in the relation (\ref{phisol}). An analysis of asymptotic behavior
at infinity yields the boundary value
\begin{equation}
\cos \theta _{0}=\frac{h}{4S(J_{\perp }-J_{z})}  \label{boundary}
\end{equation}
from which one obtain $\theta _{0}=\frac{\pi }{2}$ at $h=0$ and there is no
static solutions at $h\neq 0$ and $J_{\perp }=J_{z}$. However, in the last
case there is a dynamical solution $\varphi =\varphi (r)+\omega t$ with the
Larmor resonance frequency.

For a zero magnetic field $h$ an asymptotic behavior at $r\rightarrow \infty
$
\[
\theta =\frac{\pi }{2}+\delta \theta
\]
may be obtained from the equation
\begin{equation}
r^{2}\frac{d^{2}\delta \theta }{dr^{2}}+r\frac{d\delta \theta }{dr}-\left( 4%
\frac{(J_{\perp }-J_{z})}{J_{z}}r^{2}-\frac{J_{\perp }}{J_{z}}q^{2}\right)
\delta \theta =0.  \label{asya}
\end{equation}
The solution is the McDonald's function $\delta \theta =K_{v}\left( \frac{r}{%
\lambda }\right) $, $v=iq\sqrt{\frac{J_{\perp }}{J_{z}}}$ with the
localization radius
\begin{equation}
\lambda =\frac{1}{2}\sqrt{\frac{J_{z}}{J_{\perp }-J_{z}}}.  \label{radius}
\end{equation}
A series expansion of the solution (\ref{BOsol}) near the point $r=a$
\begin{equation}
\theta (r)\approx \theta (a)+\theta ^{^{\prime }}(a)\;r.  \label{asyazero}
\end{equation}
A numerical calculation of the Eqs. (\ref{firstani}, \ref{secondani}) is
made by the shooting method and an example of the nodal OP soliton is
presented in Fig. 4a. The phase portrait of possible behavior of the solutions presented in Fig.
5. An arbitrary choice of the derivative $\theta ^{^{\prime }}(a)$ results
in energetically unfavorable oscillating solutions above or below $\theta =%
\frac{\pi }{2}$ (left and right limit cycles) and only an unique choice of
the derivative gives the soliton. Like in the case of the conventional OP
with an axial symmetry an energy has a logarithmic divergence with a growing
of the soliton size.

Finally, we point out that an inclusion of an external magnetic field $h$
changes the asymptotic behavior of the soliton at infinity from the
exponential into the power decay
\begin{equation}
\delta \theta \approx -\frac{1}{4}\frac{J_{\bot }}{J_{\bot }-J_{z}}\frac{%
\cos \theta _{0}}{\sin ^{5}\theta _{0}}\frac{q^{2}}{r^{2}},  \label{THapr}
\end{equation}
where $\theta _{0}$ is given by (\ref{boundary}). This is a counterpart of
conventional TH solitons (Fig. 4b).

In conclusion, the class of the nodal stationary states in the 2D Heisenberg
model is investigated. The scale-invariant solution, a counterpart of BP
soliton, is found. An account of in-plane exchange anisotropy yields the
analogues of out-of-plane and Takeno-Homma solitons, however, in contrary to
the conventional solutions, those which are considered in the paper have no
axial symmetry.

%\acknowledgments
This work was partly supported by the grant NREC-005 of US CRDF (Civilian
Research \& Development Foundation), INTAS\ grant (Project N 01-0654), by
the grant ''Russian Universities'' ( UR.01.01.005).

\newpage

%Figure caption.

\begin{figure}[tbp]
\vspace{0.5mm}
\caption{The types of vector field singular points on the sphere: center
(a), focus (b), node (c).}
\end{figure}

\begin{figure}[tbp]
\vspace{0.5mm}
\caption{The space magnetization distribution in the "target". The
logarithmic scale is used. The inset shows a radial section of the plot.}
\end{figure}

\begin{figure}[tbp]
\vspace{0.5mm}
\caption{In-plane spin arragement of a nodal singularity. The dotted circle
displays spin directions on the equal distances from the center.}
\end{figure}

\begin{figure}[tbp]
\vspace{0.5mm}
\caption{The $\theta (r)$ dependence in the nodal out-of-plane (a) and
Takeno-Homma (b) soliton.}
\end{figure}

\begin{figure}[tbp]
\vspace{0.5mm}
\caption{The phase trajectories of possible solutions of Eq. (\ref{firstani}).}
\end{figure}

\end{document}